\documentclass[twocolumn,aps,prl,graphics]{revtex4}
\usepackage{graphicx}
\usepackage{amsmath}
\begin{document}

\title{Coupled ion - nanomechanical systems}
\author{L. Tian$^{1,2}$ and P. Zoller$^{1,2}$}
\affiliation{$^{1}$ Institute for Theoretical Physics, University of Innsbruck, 6020
Innsbruck, Austria \\
$^{2}$ Institute for Quantum Optics and Quantum Information of the Austrian
Academy of Sciences, 6020 Innsbruck, Austria}
\date{\today }

\begin{abstract}
We study ions in a nanotrap, where the electrodes are nanomechanical
resonantors. The ions play the role of a quantum optical system which acts
as a probe and control, and allows entanglement with or between
nanomechanical resonators.
\end{abstract}

\maketitle

Laser manipulated trapped ions are one of the prime examples of a quantum
system, where control of coherent quantum dynamics, state preparation and
measurement are achieved in the laboratory, while decoherence due to
coupling with the environment is strongly suppressed\cite{PhysToday2004}.
These achievements are illustrated by recent progresses in developing ion
traps for quantum computing and high precision measurements\cite%
{ion_trap_exp}. A key step in the future will be the realization and
integration of ion traps with micro-fabricated nanostructures, such as
segmented traps and on-chip ion traps with strong confinement\cite%
{large_scale_ion_trap}. As a new aspect, this opens the possibilities of
developing trapped ions as a quantum optical system which acts as a probe
and control, and allows entanglement with or between the quantum degrees of
freedom of mesoscopic systems\cite{Wineland_ion_trap,Schwab_Cleland_exp} -
while raising interesting questions of decoherence in a solid state
environment.

In this paper we study ions in a mesoscopic Paul trap, where suspended
nanomechanical resonators\cite{nanowire_derHeer_Schoneburg_nanotube} play
the role of tiny trap electrodes, and act as high-$Q$ nanomechanical
resonators with their own quantum degrees of freedom. Below we develop a
model of the trapped ions coupled with the flexural modes of these
nanomechanical electrodes. In particular, we investigate the possibility of
manipulation, preparation and measurement of quantum states of the flexural
modes via the laser driven ion in the limit where the trapping frequency of
the ion is resonant with the frequency of the nanomechanical oscillator.
This setup can be generalized to ion trap and nanoelectrode arrays. Another
application is quantum computing, where mesoscopic traps not only promise
very strong confinement and an associated speed up of two-qubit quantum
gates, but coupling via the nanomechanical electrodes offers new ways of
entangling internal states of ions. We also study the decoherence mechanism
for the trapped ions due to the nanomechanical and electrical couplings
which introduce quantum Brownian motion of the electrodes and limit the
ability of manipulating the mechanical modes via the ion.

\begin{figure}[tbh]
\includegraphics[width=6cm]{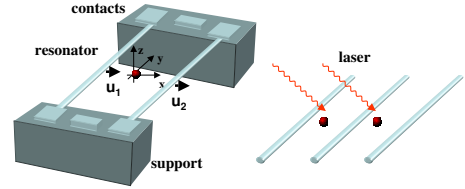}
\caption{Setup. Left: ion trap with electrodes made of nanomechanical
resonators suspended above ground and biased with ac voltage; right: arrays
of nanomechanical resonators and laser manipulation of the trapped ions.}
\label{figure_1}
\end{figure}

\textit{Model for ions coupled to nanomechanical electrodes:} We consider a
schematic setup as illustrated in Fig.~\ref{figure_1}, where an ion is
trapped between two parallel suspended electrodes represented by nanowires
or nanotubes\cite{nanowire_derHeer_Schoneburg_nanotube}. When a gate voltage
$V(t)=V_{0}\cos{\omega _{ac}t}$ with frequency $\omega _{ac}$ is applied to
the electrodes via Ohmic contacts, the charge on the resonator oscillates
with time and leads to oscillating forces on the ion as well as between the
resonators. Averaging over the fast driving frequency results in an
effectively harmonic trapping potential for the ion centered between the
electrodes. According to Fig.~\ref{figure_1} the resonators of radii $r_{0}$
are separated by a distance $2d_{0}$, much smaller than their length $2L_{0}$%
, and located a distance $h_{0}$ ($\leq L_{0}$) above the ground plane. As a
consequence, there will be tight trapping of the motion of the ion along the
$x$-axis (see Fig.~\ref{figure_1}) which couples to the high-$Q$ flexural
(bending) modes of the electrodes, compared to looser confinement in
orthogonal directions.

In Euler-Bernoulli theory\cite{Cleland_Roukes_noise} the equation of motion
for the (small) displacement $u_{i}(y)$ of the flexural modes of the $i$-th
resonator ($i=1,2$) is given by $\partial ^{2}u_{i}/\partial t^{2}+\left(
EI_{2}/\rho \right) \partial ^{4}u_{i}/\partial y^{4}=0$, where $E$ is
Young's modulus, $I_{2}$ the moment of inertia, and $\rho $ the mass density
of the electrodes\cite%
{Schwab_Cleland_exp,nanowire_derHeer_Schoneburg_nanotube}. The $n $-th
secular mode with the eigenfunction $u_{in}(y)$ has frequency $\omega _{in}=%
\sqrt{EI_{2}/\rho }q_{in}^{2}$ with wave vector $q_{in}$. We can expand the
displacement of the electrode as $u_{i}\left( y\right)
=\sum_{n}X_{in}u_{in}(y)$ where $X_{in}$ when quantized as $\hat{X}_{in}$
takes the role of a ''position'' operator with a conjugate ''momentum''
operator $\hat{P}_{in}$. The lowest flexural mode, also called the
fundamental mode, has the secular function $u_{in=1}(y)=a_{0}\left[ \cos {%
q_{i1}}\cosh {q_{i1}L}_{0}-\cosh {q_{i1}y}\cos {q_{i1}L}_{0}\right] $ where $%
a_{0}$ is the normalization coefficient and $q_{i1}\approx {(2\pi
/L_{0})} 0.376 $. Note the flexural modes have quadratic
dependence on the wave vector. Other acoustic phonons at long wave
length are the longitudinal modes and the torsional modes which
have much higher frequencies than flexural modes. For nanowires
with the length of the order of $\mathrm{\mu m} $ the secular
frequencies are in the range of tens to a few hundred MHz with
quality factors $Q \ge 10^{4}$\cite{Schwab_Cleland_exp}. For Carbon nanotube %
\cite{nanowire_derHeer_Schoneburg_nanotube,Dresselhaus_books}, the Young's
Modulus is $E=1\,\mathrm{TPa}$; with radius $r_{0}=2\,\mathrm{nm}$ and $%
2L_{0}=500\,\mathrm{nm}$ the fundamental mode has a frequency of $\omega
_{b}=1\,\mathrm{\ GHz}.$

The vibration of two suspended resonators $i=1,2$ with mass $M_{i}$ and
eigenmodes $\omega _{in}$ is described by the coupled parametric oscillator
Hamiltonian
\begin{equation}
\displaystyle H_{v}\!=\!\sum_{i,n}\left( \frac{\hat{P}_{in}^{2}}{2M_{i}}\!+\!%
\frac{M_{i}\omega _{in}^{2}}{2}\hat{X}_{in}^{2}\right) \!+\!\sum_{m,n}\kappa
_{mn}(t)\hat{X}_{1m}\hat{X}_{2n}  \label{Hv}
\end{equation}%
where the first term is the elastic energy and the second is a micromotion
of the electrodes due to the electrostatic interaction between the
oscillating charges on the resonators with $ \kappa _{mn}(t)=\kappa
_{mn}\cos ^{2}\left( \omega _{ac}t\right)$. Explicit expression for the
coefficients $\kappa _{mn}$ can be derived from an expansion of the
electrostatic energy for \emph{given} applied voltages $V_{i}(\equiv V(t))$,
$E_{m}=-\frac{1}{2}\sum_{i,j}C_{ij}(\{\hat{X}_{in}\})V_{i}V_{j}$, with $%
C_{ij}$ capacitances of the resonators as functions of the displacements $%
\hat{X}_{in}$. For an array of electrodes, as in Fig.~\ref{figure_1}, these
coupling terms may (partially) compensate each other due to symmetry
considerations.

To derive the total Hamiltonian of the ion coupled to the electrodes we must
consider the Coulomb interaction between the ion and the charge distribution
on the resonators. The charge distribution on the $i$-th electrodes includes
an (essentially uniform) charge density $\rho _{iv}\left( t\right) =\rho
_{iv}\cos \omega _{ac}t$ induced by the time dependent external voltage, and
an induced charge distribution $\rho _{iq}$ due to the presence of the ion
(``image''charge). As mentioned above, we only consider the high frequency
motion of the ion along the $x$-axis which is decoupled from the motion
along $y$ and $z$. The interaction energy between the charge of the ion $%
q_{0}$ and the total charge density on the electrodes has the form $%
\sum_{i}\int ds_{i}\left( \rho _{iv}\left( t\right) +\rho _{iq}\right)
q_{0}/4\pi \epsilon _{0}\left| R_{x}\hat{e}_{x}+y\hat{e}_{y}\right| $ with $%
R_{x}=\hat{x}+(-1)^{i+1}d_{0}-u_{i}\left( y\right) $, where we integrate
along the resonator and the denominator involves the distance between the
ion at position $\hat{x}\hat{e}_{x}$ (compare Fig.~\ref{figure_1}) and the
electrode with displacement $u_{i}\left( y\right) \hat{e}_{x}$. For $\max
\left( \left| u_{i}\left( y\right) \right| \right) \ll 2L_{0}$ the
assumption of a constant charge distribution remains a good approximation.
The interaction of the ion with the ``image charge''\ is well approximated
by $-q_{0}^{2}/4\pi \epsilon _{0}\left| R_{x}\right| \ln {(}\left| {R_{x}}%
\right| {/r_{0})}$\cite{image_charge_2002} and generates a small correction
to the trapping and the coupling energy. An additional effect here is the
modification of $\rho _{v}$ due to the small voltage of the image charge to
the ground, which can be neglected for $h_{0}\gg d_{0}$.

Second order expansion in $\hat{x}$\ and $\hat{X}_{ij}$ gives a harmonic
coupling between the ion and the electrodes. The total Hamiltonian $H_{%
\mathrm{tot}}=H_{v}+H_{\mathtt{ion}}$ with $H_{\mathrm{ion}}$ the
Hamiltonian for the ion\ of mass $m$ coupled to the electrodes
\begin{equation}
\displaystyle H_{\mathrm{ion}}=\frac{\hat{p}^{2}}{2m}+\frac{q_{0}(\widetilde{%
V}_{0}(t)+\widetilde{V}_{q})}{2d_{0}^{2}}\hat{x}^{2}+\sum_{in}g_{in}(t)\hat{X%
}_{in}\hat{x}  \label{Hion}
\end{equation}%
which includes the trapped ion Hamiltonian Hamiltonian with $\widetilde{V}%
_{0}(t)=2V_{ac}\cos {\omega _{ac}t}/\ln {(2h}_{0}\sqrt{h_{0}^{2}+d_{0}^{2}}{%
/r}_{0}d_{0}{)}$ modified by a (small) image charge contribution, $%
\widetilde{V}_{q}=-\alpha _{g}q_{0}/2\pi \epsilon _{0}d_{0}\ln {(d_{0}/r_{0})%
}$ (with $\alpha _{g}\approx 1$). The last term is the ion-electrode
coupling with $g_{in}(t)=g_{in}\cos {\omega _{ac}t}$ and
\begin{equation*}
\displaystyle g_{in}=\int_{i}dy\frac{q_{0}\rho
_{iv}(t)(y^{2}-2d_{0}^{2})u_{in}(y)}{4\pi \epsilon
_{0}(d_{0}^{2}+y^{2})^{5/2}}+\frac{q_{0}\widetilde{V}_{q}u_{in}\left(
0\right) }{2d_{0}^{2}\ln {(d_{0}/r_{0})}}
\end{equation*}%
where again the image charge gives a small modification. Note a displacement
force $\sum g_{in}^{\left( 1\right) }\cos \left( \omega _{ac}t\right) \hat{X}%
_{in}$ and an electrode-electrode interaction $\sum h_{i,mn}^{\left(
2\right) }\cos \left( \omega _{ac}t\right) \hat{X}_{im}\hat{X}_{in}$ induced
by the static ion, which can be derived by expanding the Coulomb
interaction, are not written in Eq.~(\ref{Hion}).

We are interested in a situation where the trap frequency is near resonant
with the lowest flexural mode $n=1$, while the driving field with frequency $%
\omega _{ac}$ is far off resonant with any of the other elastic eigenmodes.
This, together with the fact that $g_{in}$ is a rapidly decreasing function
of $n$, justifies the single mode approximation for the electrodes: $H_{%
\mathrm{ion}}=\hat{p}^{2}/2m+q_{0}\widetilde{V}_{0}\left( t\right) \left(
\hat{x}-u_{1}(0)\hat{X}_{p}\right) ^{2}/2d_{0}^{2} $, where $\hat{X}%
_{p}=(\alpha _{1}\hat{X}_{11}+\alpha _{2}\hat{X}_{21})/2$ and for symmetric
electrodes and $d_{0}/L_{0}\leq 1$, $\alpha _{1}=\alpha _{2}\approx 1$. This
Hamiltonian has the simple interpretation of a parametric oscillator where
the trap center located at the center-of-mass (COM) displacement $\hat{X}_{p}
$, providing a bilinear coupling of the ion to the electrodes.

As a final step, we adiabatically eliminate the (micro-)motion at the
parametric drive frequency $\omega _{ac}$. We illustrate this for the case
of two symmetric electrodes with identical eigenmode frequencies and masses,
so that the ion only couples to the COM mode $\hat{X}_{p}$. We obtain for
the effective Hamiltonian with coupling in the rotating wave approximation%
\begin{equation}
\displaystyle H_{\mathrm{eff}}=\hbar \omega _{\nu }\hat{a}^{\dagger }\hat{a}%
+\hbar \omega _{b}\hat{b}^{\dagger }\hat{b}+(i\lambda \hat{b}^{\dagger }%
\hat{a}+\mathrm{h.c.})  \label{Heff}
\end{equation}%
where $\hat{b}$ ($\hat{b}^{\dag }$) are the lowering (raising) operators of
the COM flexural mode with mass $M_{p}=2M_{i}$; and $\hat{a}$ ($\hat{a}%
^{\dag }$) operators for the harmonically trapped ion with frequency $\omega
_{\nu }=q_{0}\widetilde{V}_{0}/\sqrt{2}d_{0}^{2}m\omega _{ac}$ and mass $m$.
Near resonance at $\omega _{\nu }\sim \omega _{b}\ll \omega _{ac}$, the
coupling is proportional to the trapping frequency while decreased by the
mass ratio $m/M_{p}$: $\lambda =\hbar \omega _{\nu }u_{1}(0)\sqrt{m/M_{p}}$.
This shows the nanomechanical resonator with lighter mass, e.g.~a single
wall carbon nanotube, will have stronger coupling with the trapped ion.
Typical parameters are $V_{0}=5\,\mathrm{V}$, $\omega _{ac}/2\pi =3\,\mathrm{%
GHz}$, $d_{0}=100\,\mathrm{nm}$, $h_{0}=10d_{0}$, $r_{0}=2\,\mathrm{nm}$, $%
m/M_{p}\approx 10^{-4}$, and we have $\omega _{\nu }/2\pi =1\,\mathrm{GHz}$.
The time dependent coupling $\kappa _{mn}\cos ^{2}\omega _{ac}t$ in Eq.~(\ref%
{Hv}) satisfies $\left| \kappa _{mn}\right| \leq (Q_{c}/q_{0})\left( \omega
_{ac}/\omega _{b}\right) m\omega _{b}^{2}\ll M_{p}\omega _{b}^{2}$ at $%
\omega _{\nu }\sim \omega _{b}$, where $Q_{c}=2L_{0}\rho _{v}$ is the charge
on the electrodes from the voltage source. For the above numbers, $%
Q_{c}\approx 50$ and hence the $\kappa _{mn}$ term, so are the $\sum
g_{in}^{(1)}$ and $h_{i,mn}^{(2)}$ terms, can be neglected due to the small
mass ratio. In general, cantilever couplings are readily accounted for by
transforming to a set of eigenmodes.

\begin{figure}[tbp]
\includegraphics[width=6.5cm]{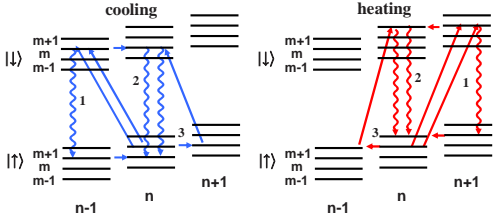}
\caption{Energy structure of the combined system of the ion and the
electrodes. Left: cooling circles; right:heating circles.}
\label{figure_2}
\end{figure}

The coupled oscillator Hamiltonian (\ref{Heff}) allows the transfer of the
motional states of the electrodes and the ion. Motional states of the ions
can be manipulated by coupling internal electronic states to a laser: $%
H_{I}=-\frac{1}{2}\delta \sigma _{z}+\Omega e^{i\eta (a+a^{\dag })}\sigma
^{+}+\mathrm{h.c.,}$ which can be added to Eq.~(\ref{Heff}). Here $\sigma $%
's are the Pauli operators of the a two-state system, $\delta $ the laser
detuning, $\Omega $ the Rabi frequency, and $\eta =k\sqrt{\hbar /2m\omega
_{\nu }}$ the Lamb-Dick parameter describing the laser recoil on the ion
motion. Based on these couplings, a complete toolbox is available on the ion
for (i) quantum state engineering, and (ii) preparation of pure states
(ground state cooling), and (iii) state measurements using the quantum jump
technique. Preparing and analyzing motional states of the resonators are
available via the transfer Hamiltonian when the coupling time $1/\lambda$ is
faster than the decoherence time.

\textit{Decoherence:} Decoherence of the nanomechanical resonator is induced
by mechanical\cite{Cleland_Roukes_noise} and electrical noise\cite%
{Blanter_Buttiker_PR}. The mechanical noise is characterized by the quality
factor $Q$ due to the coupling with environments, such as the noise in the
support of the electrodes and the surface of the electrodes. Electrical noise%
\cite{Blanter_Buttiker_PR} represents the voltage fluctuations on the
electrodes, including the noise of the resistances in the circuit, shot
noise and low frequency noise ($1/f$ noise). The shot noise is negligible
for a ballistic electrode, i.e. single wall Carbon nanotube\cite%
{Blanter_Buttiker_PR}. The low frequency noise is due to charge fluctuations
in the environment and the imperfection or dirt on the electrodes, and is
negligible at GHz frequency.

Let $\delta \widetilde{v}_{i}\left( t\right) $ be the noise on the
electrodes defined the same way as $\widetilde{V}_{0}$ with $i=1,2$. The
voltage noise introduces: the parametric noise $q_{0}\delta \widetilde{v}%
_{p}\left( t\right) \left( \hat{x}-u_{1}\hat{X}_{p}\right) ^{2}/2d_{0}^{2}$
with the spectrum $\delta \widetilde{v}_{p}^{2}\left( \omega \right) \approx
\sum \delta \widetilde{v}_{i}^{2}\left( \omega \right) /4$, and the linear
noise $q_{0}\delta \widetilde{v}_{m}\left( t\right) \left( \hat{x}-u_{1}\hat{%
X}_{p}\right) /2d_{0}$ with $\delta \widetilde{v}_{m}^{2}=\beta _{g}^{2}\sum
\delta \widetilde{v}_{i}^{2}\left( \omega \right) /4$ and $\beta _{g}\approx
2\ln \left( 2h_{0}/r_{0}\right) /\ln \left( 2d_{0}/r_{0}\right) $, when $%
\delta \widetilde{v}_{1,2}\left( t\right) $ are not correlated. Due to $%
m/M_{p}\ll 1$, the $\hat{X}_{p}^{2}$, $\hat{x}\hat{X}_{p}$ and $\hat{X}_{p}$
terms are a factor of $m/M_{p}$ or $\left( m/M_{p}\right) ^{2}$ smaller than
the $\hat{x}^{2}$ and $\hat{x}$ terms, so we only consider the $\hat{x}^{2}$
and $\hat{x}$ terms. The parametric noise has a spectral density $\gamma
_{1}(\omega )=\left( \delta x_{0}/d_{0}\right) ^{4}q_{0}^{2}\delta
\widetilde{v}_{p}^{2}\left( \omega \right) /4$. Due to the small ratio $%
\delta x_{0}/d_{0}$, $\gamma _{1}=0.5\mathrm{Hz}$ for the contact resistance
at zero temperature and can be neglected. Here we have $R_{c}=h/2N_{\perp
}e^{2}$ with $N_{\perp }$ the number of conducting channels and $N_{\perp
}=4 $ for Carbon nanotube, and $\delta \widetilde{v}_{p}^{2}\left( \omega
\right) =R_{c}\omega /2$ at zero temperature. The linear term has a spectral
density $\gamma _{m}(\omega )=\left( \delta x_{0}/d_{0}\right)
^{2}q_{0}^{2}\delta \widetilde{v}_{m}^{2}\left( \omega \right) /4$, with $%
\delta \widetilde{v}_{m}^{2}\left( \omega \right) =R_{c}\omega /2$ for
uncorrelated noise and $\gamma _{m}(\omega _{\nu })=0.5\mathrm{MHz}$. Note
that due to the $\left( \delta x_{0}/d_{0}\right) ^{2}$ factor, this term is
stronger than the parametric noise. The linear noise can be avoided when the
two electrodes are made of one piece of metallic wire connecting to only one
contact. Now $\delta \widetilde{v}_{1}=\delta \widetilde{v}_{2}$, and $%
\gamma _{m}\sim 0$ as well as $\delta \widetilde{v}_{m}^{2}\sim 0$. Thus the
condition of a decoherence time longer than the coupling time of the ion to
the resonator can be met.

\textit{Resonator Cooling:} The energy structure of the system is shown in
Fig.\ref{figure_2}, where the state $\left| s,n,m\right\rangle $ corresponds
to the internal state at $s$, the motion of the ion at the number state $n$,
and the vibrational mode at the number state $m$. The states $\left|
s,n,m\right\rangle $, $\left| s,n-1,m+1\right\rangle $ and $\left|
s,n+1,m-1\right\rangle $ are connected by the coupling $\lambda $.
Eliminating the internal degrees of freedom of the ions, the master equation
in the interaction picture of the ion-resonator motion becomes
\begin{equation}
\begin{array}{r}
\displaystyle\frac{\partial \rho }{\partial t}=-i\left[ \lambda \hat{a}%
^{\dagger }\hat{b}+\mathrm{h.c.},\rho \right] +\mathcal{L}^{d}\left( \hat{a}%
,\gamma _{\omega }^{\hat{a}},n_{f}\right) \rho \\
\displaystyle+\mathcal{L}^{d}\left( \hat{b},\frac{\omega _{B}}{Q}%
,n_{B}\right) \rho +\mathcal{L}^{d}\left( \hat{a},\gamma _{m},n_{B}\right)
\rho%
\end{array}
\label{rho}
\end{equation}%
where we use the notation in \cite{Cirac_Blatt_Zoller_Phillips}
\begin{equation*}
\begin{array}{r}
\displaystyle\mathcal{L}^{d}\left( A,\gamma _{\omega }^{A},n_{\omega
}\right) \rho =\frac{1}{2}\gamma _{\omega }^{A}n_{\omega }\left( 2A^{\dag
}\rho A-AA^{\dag }\rho -\rho AA^{\dag }\right) \\[2mm]
\displaystyle+\frac{1}{2}\gamma _{\omega }^{A}\left( n_{\omega }+1\right)
\left( 2A\rho A^{\dag }-A^{\dag }A\rho -\rho A^{\dag }A\right)%
\end{array}%
\end{equation*}%
for the Liouvillian operators with the system operator $A$, noise spectrum $%
\gamma _{\omega }^{A},$ and a phonon number $n_{\omega }$. In Eq.~(\ref{rho}%
) laser cooling of the ion contributes the $\mathcal{L}^{d}\left( \hat{a}%
,\gamma _{\omega }^{\hat{a}},n_{f}\right) $ term, where $n_{f}$ is the final
phonon number of the ion in the absence any additional heating mechanisms
(i.e. $n_{f}\approx 0$) and cooling rate $\gamma _{\omega }^{\hat{a}}=\eta
^{2}\Omega _{r}^{2}/\Gamma _{e}$ when $\Gamma _{e}>\eta \Omega _{r}$ is the
rate of laser cooling. The linear voltage noise contributes the $\mathcal{L}%
^{d}\left( \hat{a},\gamma _{m},n_{B}\right) \ $term. The resonator damping
and heating contributes $\mathcal{L}^{d}\left( \hat{b},\gamma _{\omega }^{%
\hat{b}},n_{B}\right) $ where $\gamma _{\omega }^{\hat{b}}=\omega _{b}/Q$
and $n_{B}=k_{B}T/\hbar \omega _{b}$ is the thermal phonon number of the
supports at frequency $\omega _{b}$.

For $n_{f}=0$, the final state phonon number of the electrodes is
\begin{equation*}
\langle \hat{b}^{\dag }\hat{b}\rangle _{f}=\frac{\displaystyle\left(
\,\Gamma _{\nu }\,\gamma _{\mathrm{eff}}^{\hat{a}}\left( \gamma _{\mathrm{eff%
}}^{\hat{a}}+\Gamma _{\nu }\right) +4\Gamma _{\nu }\lambda ^{2}\right)
n_{B}+4\,\,\gamma _{\mathrm{eff}}^{\hat{a}}\,{\lambda }^{2}n_{f}^{\mathrm{eff%
}}}{\displaystyle{\left( \Gamma _{\nu }+\gamma _{\mathrm{eff}}^{\hat{a}%
}\right) }\left( \gamma _{\mathrm{eff}}^{\hat{a}}\Gamma _{\nu }+4\lambda
^{2}\right) }
\end{equation*}%
with $\Gamma _{v}=\omega _{b}/Q$. The effective cooling rate is $\gamma _{%
\mathrm{eff}}^{\hat{a}}=$ $\gamma _{\omega }^{\hat{a}}+\gamma _{m}$
including the voltage noise with the final phonon number of the ion $n_{f}^{%
\mathrm{eff}}=n_{B}\gamma _{m}/\gamma _{\mathrm{eff}}^{\hat{a}}$. When $%
\gamma _{m}\sim \gamma _{\mathrm{eff}}^{\hat{a}}$ the ion can not be cooled
and hence the electrodes. As a necessary requirement we need $\gamma
_{m}\sim 0$, i.e. the two electrodes in one piece (see above). In this case,
when $\gamma _{\omega }^{\hat{a}}=2\lambda $ the stationary phonon number is
minimized: $\langle \hat{b}^{\dag }\hat{b}\rangle _{\min }=\left(
k_{B}T/Q\right) (4\lambda +\Gamma _{\nu })/\left( 2\lambda +\Gamma _{\nu
}\right) ^{2}$. With our parameters $\Gamma _{e}=5\,\mathrm{MHz}$, $\Omega
_{r}=300\,\mathrm{MHz}$, and $\gamma _{\omega }^{\hat{a}}=2\mathrm{MHz}$, we
have $\langle \hat{b}^{\dag }\hat{b}\rangle _{f}\approx 0.5$ with $Q=10^{5}$
and $\mathrm{T}=4\mathrm{K}$.

For the high trapping frequencies, the Lamb-Dicke\ parameter is small with $%
\eta \sim 0.01$ and as a result $\gamma _{\omega }^{\hat{a}}<\lambda $. The
ion cooling rate becomes the bottleneck for the cooling process. The problem
can be avoided by trapping the ion at low frequency $\omega _{\nu }\ll
\omega _{b}\sim \omega _{ac}$ (and thus large $\eta $). The ac driving field
provides a parametric up-conversion of the ion to the resonator phonons,
i.e. $\lambda \longrightarrow \hbar \omega _{\nu }u_{1}\sqrt{m\omega
_{ac}^{2}/M_{p}\omega _{b}\omega _{\nu }}e^{-i\omega _{ac}t}$ in Eq. (\ref%
{Heff}).

\textit{Entanglement generation: }Quantum state engineering of ion can be
used to generate entangled states of the resonators via the coupling Eq.~(%
\ref{Heff}). For example, let us discuss the generation of $\left| \psi
_{1},\chi _{2}\right\rangle +\left| \chi _{1},\psi _{2}\right\rangle $ for
two resonators where $\left| \psi _{i}\right\rangle $ and $\left| \chi
_{i}\right\rangle $ are arbitrary states of the electrode $i=1,2$. We choose
the electrodes to have different fundamental frequency $\omega _{b1}\neq
\omega _{b2}\,$ and coupling with the ion $\lambda _{i}$. The initial state $%
\left( \left| \uparrow ,\psi _{x}\right\rangle +\left| \downarrow ,\chi
_{x}\right\rangle \right) \left| 0_{1},0_{2}\right\rangle $ with $x$ for the
motion of the ion is prepared using standard protocols of
the ion trap qubits\cite{Eberly_Law_Cirac_Zoller}, where the motional and the
internal state of the ion are entangled. In a first step we tune the ion to
the first resonator resonance $\omega _{\nu }=\omega _{b1}$ for a duration
of $\pi /2\lambda _{1}$, which results in the swap $\left|
n_{x},m_{1}\right\rangle \rightarrow \left| m_{x},n_{1}\right\rangle $, and
the state is now $\left( \left| \uparrow ,0_{x},\psi _{1}\right\rangle
+\left| \downarrow ,0_{x},\chi _{1}\right\rangle \right) \left|
0_{2}\right\rangle $. Now we prepare the state $\left( \left| \uparrow ,\chi
_{x},\psi _{1}\right\rangle +\left| \downarrow ,\psi _{x},\chi
_{1}\right\rangle \right) \left| 0_{2}\right\rangle $ via a third internal
state in the ion\cite{Eberly_Law_Cirac_Zoller}. Then, we tune the trapping
frequency to $\omega _{\nu }=\omega _{b2}$ for a duration of $\pi /2\lambda
_{2}$, which results in $\left| n_{x},m_{2}\right\rangle \rightarrow \left|
m_{x},n_{2}\right\rangle $, and the state is $\left( \left| \uparrow
,0_{x},\psi _{1},\chi _{2}\right\rangle +\left| \downarrow ,0_{x},\chi
_{1},\psi _{2}\right\rangle \right) $. Now we rotate the internal state by a $%
\pi /2$ pulse: $\left| \uparrow \right\rangle \rightarrow \left| \uparrow
+\downarrow \right\rangle $ and $\left| \downarrow \right\rangle \rightarrow
\left| \uparrow -\downarrow \right\rangle $; then detect the internal state.
The detection generates states $\left| \psi _{1},\chi _{2}\right\rangle \pm
\left| \chi _{1},\psi _{2}\right\rangle $ depending on the detected internal
states; and entanglement is transferred from the ion to the resonators. This
method can be applied to generate for example $\left|
0_{1},n_{2}\right\rangle +\left| n_{1},0_{2}\right\rangle $ as entangled
state of the two electrode modes. The entanglement can be achieved on a time
scale of $\pi /\lambda _{1,2}\approx 50\mathrm{n}\sec $. Given a decoherence
time of $1\mathrm{MHz}$, a fidelity exceed $0.9$ can be achieved.

\textit{Quantum Computing:} Mesoscopic traps promise trap
frequencies significantly higher than those of present ion traps
and an associated speed up of two-qubit gates in ion trap quantum
computing (Fig. \ref{figure_1}). The standard 2-qubit protocols
based on, for example, entanglement via a phonon data bus or a
push gate are readily adapted to the present case. In the second
case, gate times of the order of nanoseconds seem possible for the
numbers above. A second possibility is entanglement via exchange
of phonons of one of the collective ion-electrode modes of the
system. Additional noise is, however introduced by the mechanical
motion. The decoherence of the ion motion when in resonance with
one mode $\omega _{\nu }=\omega _{i1}$ is $\tau
_{res}^{-1}=k_{B}T/Q$ which is $1\mathrm{MHz}$ with $Q=10^{5}$ and
assuming the same quality factor for two electrodes. When off
resonance $\left| \omega _{i1}-\omega _{\nu }\right| >k_{B}T/Q$,
the decoherence rate is
\begin{equation}
\displaystyle\Gamma _{m}=\frac{k_{B}T}{Q}\frac{m}{M_{p}}\sum_{i,n}\frac{%
u_{i_{b}}^{2}(0)\omega _{\nu }^{3}\omega _{i_{n}}}{(\omega _{\nu
}^{2}-\omega _{in}^{2})^{2}+4\omega _{in}^{2}\left( \frac{k_{B}T}{Q\hbar }%
\right) ^{2}}
\end{equation}%
so that the decoherence rate is $\Gamma _{m}\approx
(8u_{1}^{2}m/9M_{p})\left( k_{B}T/Q\right) \sim 2\pi \cdot 100\,\mathrm{Hz}$
with $\omega _{i1}=500\,\mathrm{MHz}$ and $\omega _{\nu }=1\,\mathrm{GHz}$.
The small mass ratio $m/M_{p}$ and the off resonance protect the ion from
decoherence. The decoherence due to the electrical noise is, dominated by $%
4\gamma _{1}\left( 0\right) =2\pi \cdot 0.1\mathrm{Hz}$ assuming a low
frequency noise $\left( q_{0}^{2}\delta \widetilde{v}_{1}^{2}\left( 0\right)
\right) ^{-1}=1\mathrm{n}\sec$. Hence, the decoherence time is much longer
in the nano trap. This shows that the smallest charge box -- single ion --
coupling with the nanomechanical resonators not only presents an effective
knob for quantum features of the flexural modes, but also increases the
speed of ion trap quantum computing by several orders of the magnitudes.

Acknowledgments: We thank M.S. Dresselhaus, B.I. Halperin, D. Leibfried,
L.S. Levitov, M.D. Lukin, A.S. S{\o}rensen, and W. Zwerger for helpful
discussions. This work is supported by the Austrian Science Foundation,
European Networks and the Institute for Quantum Information.


\vskip -5mm

\begin{thebibliography}{99}
\bibitem{PhysToday2004} J. I. Cirac and P. Zoller, Phys. Today March vol.%
\textbf{57}, 38 (2004).

\bibitem{ion_trap_exp} M. D. Barrett \textit{et al.}, Nature \textbf{429},
737 (2004); M. Riebe \textit{et al.}, Nature \textbf{429}, 734 (2004)

\bibitem{large_scale_ion_trap} D. Kielpinski \textit{et al.}, Nature \textbf{%
417}, 709 (2002); J. I. Cirac and P. Zoller, Nature \textbf{404}, 579 (2000).

\bibitem{Wineland_ion_trap} D.J. Wineland \textit{et al.}, J. Res. Natl.
Inst. Stand. Technol. \textbf{103}, 259 (1998), see sect. 6.4, pp319; L.
Tian and \textit{et al.}, Phys. Rev. Lett. \textbf{92}, (2004) 247902; A. S.
S{\o}rensen \textit{et al.}, Phys. Rev. Lett. \textbf{92}, 063601 (2004).

\bibitem{Schwab_Cleland_exp} M.D. LaHaye \textit{et al.}, Science \textbf{304%
}, 74 (2004); R.G. ~Knobel and A.N. Cleland, Nature \textbf{424}, 291 (2003).

\bibitem{nanowire_derHeer_Schoneburg_nanotube} A. Husain \textit{et al.},
Appl. Phys. Lett. \textbf{83}, 1240 (2003); P.~Poncharal \textit{et al.},
Science \textbf{283}, 1513 (1999); B.~Babic \textit{et al.}, Nano Lett.
\textbf{3}, 1577 (2003).

\bibitem{Cleland_Roukes_noise} A.N. Cleland and M.L. Roukes, J. Appl. Phys.
\textbf{92}, 2758 (2002).

\bibitem{Dresselhaus_books} R. Saito \textit{et al.}, \emph{Physical
Properties of Carbon Nanotubes}, Imperial College Press, London (1998); MRS
Bulletin \textbf{29}, \emph{Advances in Carbon Nanotubes}, April 2004.

\bibitem{image_charge_2002} B.E. Granger \textit{et al.}, Phys. Rev. Lett.
\textbf{89}, 135506 (2002).

\bibitem{Blanter_Buttiker_PR} Ya. M. Blanter and M. B\"{u}ttiker, Phys. Rep.
\textbf{336}, 1 (2000).

\bibitem{Cirac_Blatt_Zoller_Phillips} J. I. Cirac \textit{et al.}, Phys.
Rev. A \textbf{46}, 2668 (1992).

\bibitem{Eberly_Law_Cirac_Zoller} C. K. Law and J. H. Eberly, Phy. Rev.
Lett. \textbf{76}, 1055 (1996); J. I. Cirac and P. Zoller, Phys. Rev. Lett.
\textbf{74}, 4091 (1995).
\end{thebibliography}
\end{document}